\documentclass[journal]{IEEEtran}
\usepackage{cite}

\ifCLASSINFOpdf
  \usepackage[pdftex]{graphicx}
\else
\fi
\usepackage[caption=false, font=footnotesize]{subfig}

\usepackage{amsmath}
\usepackage{amsmath,mleftright}
\usepackage{amssymb}
\usepackage{amsthm}
\usepackage{algorithm}
\usepackage{algorithmic}

\usepackage{array,multirow}

\begin{document}
%
\title{Enhanced Wi-Fi RTT Ranging:\\A Sensor-Aided Learning Approach}

%
%
%

\author{Jeongsik~Choi,~\IEEEmembership{Member,~IEEE}
\thanks{The author was with the Intel Labs, Intel Corporation, Santa Clara, CA 95054, USA. He is now with the School of Electronics Engineering, Kyungpook National University, Daegu 41566, South Korea (e-mail: jeongsik.choi@knu.ac.kr).}
}

\maketitle

\begin{abstract}
The fine timing measurement (FTM) protocol is designed to determine precise ranging between Wi-Fi devices using round-trip time (RTT) measurements.
However, the multipath propagation of radio waves generates inaccurate timing information, degrading the ranging performance.
In this study, we use a neural network (NN) to adaptively learn the unique measurement patterns observed at different indoor environments and produce enhanced ranging outputs from raw FTM measurements.
Moreover, the NN is trained based on an unsupervised learning framework, using the naturally accumulated sensor data acquired from users accessing location services. 
Therefore, the effort involved in collecting training data is significantly minimized.
The experimental results verified that the collection of unlabeled data for a short duration is sufficient to learn the pattern in raw FTM measurements and produce improved ranging results.
The proposed method reduced the ranging errors in raw distance measurements and well-calibrated ranging results requiring the collection of ground truth data by 47--50\% and 17--29\%, respectively.
Consequently, positioning error reduced by 17--30\% compared to the result with well-calibrated ranging.
\end{abstract}


%
\IEEEpeerreviewmaketitle

\section{Introduction}

\IEEEPARstart{L}{ocation} information is essential for various mobile applications, such as positioning, navigation, vehicle/asset tracking, and vehicle to vehicle communications~\cite{8581405, 8672615, 9136596}. 
Typically, the global navigation satellite system (GNSS) plays a key role in locating mobile devices outdoors.
However, a severe penetration loss due to the outer walls makes a satellite-based positioning solution unstable in indoor environments.
Therefore, several approaches have been introduced to enable indoor positioning capability~\cite{correa_sen_17,8409950, 9253514}.

In many cases, Wi-Fi is a primary component used to locate mobile devices.
Numerous access points (APs) are deployed in most indoor sites, and mobile devices measure the signal strength conveniently from nearby APs by monitoring beacon broadcasts.
The received signal strength (RSS) measurements are used to either estimate the distance from each AP for range-based positioning~\cite{Wang2003AnIW,5425237,5558044,5766644,7736965, 7935650, 9530676} or generate a database for fingerprinting-based positioning approaches~\cite{832252,1047316,horus05,Liu2012,8007254}.
Additionally, more diverse measurements have been used to locate Wi-Fi devices, such as channel state information (CSI)~\cite{7438932, 8027020, 8166758, 8187642, 8485917, 9446615} and angle of arrival (AoA)~\cite{10.5555/2482626.2482635, 10.1145/2785956.2787487, Ahmed2018}.

The fine timing measurement (FTM) protocol, defined in the IEEE 802.11-2016 standard, considers round-trip time (RTT) measurement between two Wi-Fi devices~\cite{7786995}.
The RTT between the devices is obtained by exchanging multiple wireless packets containing timing information, such as time of departure (ToD) and time of arrival (ToA) of packets; the obtained value is converted to a distance. 
Moreover, as the standard defines all timing information in nanoseconds, a sub-meter level ranging accuracy can be achieved theoretically.

Owing to the availability of FTM-capable devices, several papers investigated the benefits of the FTM protocol.
Initially, Banin~\emph{et al.} demonstrated the positioning accuracy~\cite{ipin16} and investigated low-level fundamentals, such as the impact of clock drift on ranging performance~\cite{8579543}.
Additionally, both ranging and positioning performances have been widely evaluated using the FTM protocol in various indoor and outdoor environments~\cite{ibrahim_mobicom_18, 8756276, 8839041, 8911824, 8911751, 8911783, 8924707, 9107223, 9110232, 9125898, 9151400, 9155464, 9217193, s20051489, s20164515, Yu_2019, han2020exploiting, app10030956}.
Certain studies investigated methods to improve the ranging accuracy using both RSS and RTT measurements~\cite{8924707, 8839041} or multiple RTT measurements obtained from different frequency bands~\cite{s20051489}.
Moreover, a high-resolution estimation technique was proposed to improve the ranging accuracy when the CSI of FTM packets is available~\cite{9155464}.
In the positioning stage, Bayesian filtering techniques have been applied to track the location of mobile devices using a series of ranging results.
In addition, inertial sensors incorporated in mobile devices can be used in this stage to improve the positioning accuracy further~\cite{8756276, 9107223, 9125898, Yu_2019, han2020exploiting}.

However, the FTM protocol is known to produce biased ranging outputs, which must be calibrated to achieve accurate ranging and positioning performances.
Furthermore, some papers indicated that the FTM protocol generates a higher number of ranging errors in non-line-of-sight (NLOS) conditions than that in line-of-sight (LOS) conditions~\cite{9155464, app10030956}.
In this situation, this work studies a machine learning-based approach to obtain precise ranging results.
To this end, we deploy a neural network (NN) that takes raw measurements reported from the FTM protocol and produces enhanced ranging outputs.
Regardless of biased raw distance measurements and different measurement patterns observed at multiple indoor sites or between LOS and NLOS conditions at the same site, the NN learns measurement patterns flexibly to improve the ranging accuracy.

To minimize the efforts involved in training data collection, we adopted the sensor-aided learning framework proposed in our previous study~\cite{choi2020sensoraided}.
As several positioning applications utilize inertial sensors in mobile devices, sensor data can be collected conveniently when users access location services. In the training stage, the collected sensor data are used to estimate the trajectory of mobile devices using various sensor fusion techniques.
For instance, the pedestrian dead reckoning (PDR) method is widely used when users walk with mobile devices~\cite{Tian2014, 6987239, 8385119}.
The estimated trajectory obtained from these sensor data can be used as a reference to train the Wi-Fi ranging module.
Thus, the proposed NN can be trained with unlabeled training data collected from multiple users.

Furthermore, the unsupervised learning feature is essential to deploy positioning solutions and investigate the characteristics of each site rapidly. Although mobile devices may initially rely on raw distance measurements from the FTM protocol to obtain the location, as soon as sufficient amount of unlabeled data are accumulated, measurement patterns observed at a specific site can be learned to produce enhanced ranging results.
In other words, the ranging module is autonomously optimized while users access location services. 
Moreover, location service providers can build databases for optimal ranging strategies at each site, enabling the utilization of pre-trained ranging modules.

The contributions of this study are summarized as follows.
\begin{enumerate}
\item[1)] The ranging performance of the FTM protocol is evaluated using the latest Wi-Fi chipset. In particular, we verify the impact of transmission bandwidths and channel conditions on the ranging accuracy, success rate, and measurement error pattern of the FTM protocol.

\item[2)] We deploy an NN that takes raw measurements from the FTM protocol to produce enhanced ranging results. The proposed ranging module is autonomously trained without human intervention as unlabeled training data are accumulated when users access location services.

\item[3)] The positioning performance is evaluated for two scenarios, wherein the Wi-Fi ranging is used independently and sensors are combined with ranging results.
Therefore, the need for sensors can be determined based on the level of location accuracy required for each application.

\end{enumerate}

The remainder of the paper is organized as follows. Section~II presents a brief overview of the FTM protocol, positioning, and sensor-fusion techniques.
The raw ranging results collected in an indoor office environment are presented in Section~III. Section~IV discusses the ranging module using an NN to produce enhanced ranging results.
The positioning performance using Wi-Fi ranging with and without sensors is presented in Section~V, followed by conclusions.
The major variables defined in this paper are summarized in Table~\ref{tab_variable_list}.

\begin{table}
    \renewcommand{\arraystretch}{1.25}
    \caption{List of Variables}
    \centering
    \begin{tabular}{cl}
    \hline
    \multicolumn{1}{c}{\bf Variable} & \multicolumn{1}{c}{\bf Description}\\
    \hline
    $\tau$  & Round trip time (RTT) between Wi-Fi devices \\
    $B$     & Number of packet exchanges for the FTM burst mode\\
    $d_{ftm}$   & Average of $B$ distances measured from the FTM protocol\\
    $s_{ftm}$   & Standard deviation of $B$ distance measurements\\
    $p_{ftm}$   & Average signal strength of $B$ received packets\\
    $N$     & Number of Wi-Fi APs \\
    $\mathbf z_n$   & Position of AP $n$ $(n=1, ..., N)$\\
    $\mathbf z$     & Position of the mobile device\\
    $\hat{\mathbf z}^{(k)}$ & Estimated position of the device at time step $k$\\
    $\mathcal R^{(k)}$ & Set of ranging results from all APs at time step $k$\\
    $\mathbf r_n^{(k)}$ & Distance and standard deviation from AP $n$ at time step $k$\\
    $\mathbf p(t)$ & Position obtained using sensors at sensor timestamp $t$\\
    $\Lambda(t)$ & Step length detected at sensor timestamp $t$\\
    $\phi(t)$ & Heading direction of the device at sensor timestamp $t$\\
    $\phi_{ref}$ & Reference heading direction\\
    $\mathbf g(\mathbf x; \Theta)$ & NN-based ranging module\\
    $\mathbf x$ & Input of the ranging module  ($\mathbf x=[d_{ftm}, s_{ftm}, p_{ftm}]^T$)\\
    $\Theta$    & Set of trainable parameters in the ranging module\\
    $\hat{d}$ & Distance output from the ranging module\\
    $\hat{s}$ & Standard deviation output from the ranging module\\
    $\hat{\mathbf z}^{(k)}(\Theta)$ & Indicate that $\hat{\mathbf z}^{(k)}$ varies depending on parameters in $\Theta$\\
    $\mathcal R^{(k)}(\Theta)$ & Indicate that $\mathcal R^{(k)}$ varies depending on parameters in $\Theta$\\
    $\mathbf p^{(k)}$ & Sampled version of $\mathbf p(t)$ at time step $k$\\
    $\mathcal L(\Theta)$ & Cost function that scores the similarity of $\hat{\mathbf z}^{(k)}(\Theta)$ and  $\mathbf p^{(k)}$\\
    \hline
    \end{tabular}
    \label{tab_variable_list}
\end{table}

\section{Preliminaries}

\subsection{FTM Protocol}

The FTM protocol determines the distance between a pair of Wi-Fi devices operating at different modes, namely the initiator and responder.
In this study, the initiator and responder refer to a mobile device and an AP, respectively.

\begin{figure}
    \centering
    \includegraphics[width=0.4\textwidth]{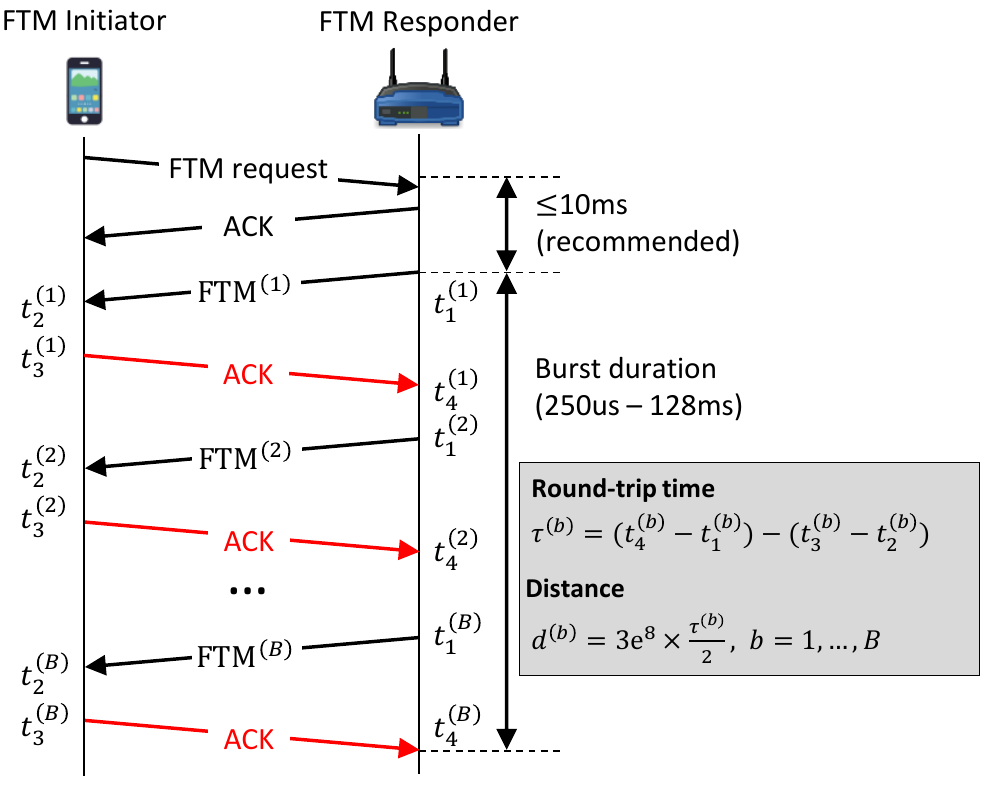}
    \caption{Signal flow of the FTM protocol.}
    \label{fig_ftm_overview}
\end{figure}

Fig.~\ref{fig_ftm_overview} illustrates the fundamental signal flow of the FTM protocol.
An initiator triggers the ranging procedure by sending an FTM request packet that contains the address of a target responder, and the target responder accepts this request by sending an acknowledgement (ACK) packet back to the initiator.
Then, the responder sends the first FTM packet in a few milliseconds and records the timestamp $t_1$ when the packet is transmitted.
As soon as the initiator receives the FTM packet, it sends an ACK packet back, which contains the ToA of the received FTM packet and ToD of the ACK packets denoted by $t_2$ and $t_3$, respectively.
Finally, the responder measures the ToA of the ACK packet as $t_4$ and computes the RTT as
\begin{equation}
    \tau = (t_4 - t_1) - (t_3 - t_2).
\end{equation}
Thus, the round-trip distance between the two devices is obtained by multiplying the value $\tau$ with the speed of light.

The part dominating the ranging error measures the ToA of the packet transmitted from the counterpart device (i.e., $t_2$ and $t_4$).
This is because the devices receive superposed multipath components, making it challenging for a relatively narrow bandwidth of Wi-Fi systems to extract the component that arrived first.
To suppress the ranging error, the FTM protocol supports the burst mode, wherein more than one pairs of FTM/ACK packets are exchanged in a single ranging request.
Fig.~\ref{fig_ftm_overview} shows that $B$ pairs of packets are exchanged. 
As each pair of packets provides an RTT measurement, $B$ distance measurements are available when a single ranging procedure completes. The FTM protocol reports the average and standard deviation of these measurements along with the RSS.

\subsection{Positioning Using Wi-Fi Ranging Results}

We consider a positioning scenario in a two-dimensional space where $N$ APs are placed at fixed positions.
The position of the device and the $n$-th AP are denoted by $\mathbf{z}=[x,y]^T$ and $\mathbf{z}_n=[x_n,y_n]^T$, respectively.
We assume that the position of every AP is given and the device performs the ranging procedure with respect to all APs at each time step.

The position of the device is obtained based on the availability of ranging results from multiple APs. 
We apply an extended Kalman filter (EKF)-based method, which yields accurate positioning performance using a series of ranging results. Because the implementation of the EKF method has been extensively investigated in the literature, this study simply defines the input and output relationship.
The details of the EKF design used in this study are presented in~\cite{8839041}.

After performing the $k$-th ranging procedure with respect to every AP, the position of the device is estimated as a function of the previously estimated position and ranging results obtained at the current time step as follows:
\begin{equation} \label{2B1}
    \hat{\mathbf z}^{(k)} = \mathbf f(\hat{\mathbf z}^{(k-1)}, \mathcal R^{(k)}),~k\geq 1,
\end{equation}
where $\hat{\mathbf z}^{(k)}$ represents the estimated position at time step $k$ and $\mathcal R^{(k)} = \{\mathbf{r}^{(k)}_n\}_{n=1}^N$ is the list of ranging results with respect to all APs.
Furthermore, the $n$-th element in the list is a vector $\mathbf{r}_n^{(k)}=[{d}^{(k)}_n, {s}^{(k)}_n]^T$ comprising an estimated distance from the $n$-th AP at time step $k$ and its standard deviation.
In this work, we process the raw FTM measurements using an NN to produce accurate ranging results for positioning.

\subsection{Trajectory Estimation Using Built-In Sensors}

Several positioning applications utilize the inertial sensors incorporated in mobile devices to estimate the trajectory.
In this study, we used the PDR method to estimate the trajectory assuming that the users walk around the site with their mobile devices~\cite{Tian2014, 6987239, 8385119}.
First, the PDR method estimates the orientation of the device with respect to a reference frame, wherein the first two axes are aligned with the x- and y-axes of the positioning space, respectively, and the final axis indicates the vertical direction.
Afterward, the local acceleration of the device is transformed into acceleration in the reference frame, and a periodic vertical movement pattern is captured to count the number of steps of a user holding the device.

Thus, the moving distance of the device is obtained by multiplying the number of detected steps with a predefined step length.
The trajectory of the device in the x-y plane is obtained by consolidating the moving distance and heading direction obtained from the estimated orientation of the device.
Therefore, the two dimensional position vector of the device at sensor time $t$ is updated as
\begin{equation} \label{2b1}
    \mathbf p(t) = \mathbf p(t-1) + \Lambda(t)
    \begin{bmatrix}
    -\sin(\phi(t)+\phi_{ref})\\ \cos(\phi(t)+\phi_{ref})
    \end{bmatrix},
\end{equation}
where $\Lambda(t)$ indicates the step length between time $t-1$ and $t$. If no step is detected in this interval, it is simply given by $\Lambda(t)=0$.
In addition, $\phi(t)$ represents the heading direction at time $t$, which is the rotation angle of the device in the x-y plane computed considering an arbitrary reference direction denoted by $\phi_{ref}$.

\section{Measurement Campaigns}

\begin{figure}
    \centering
    \includegraphics[width=0.4\textwidth]{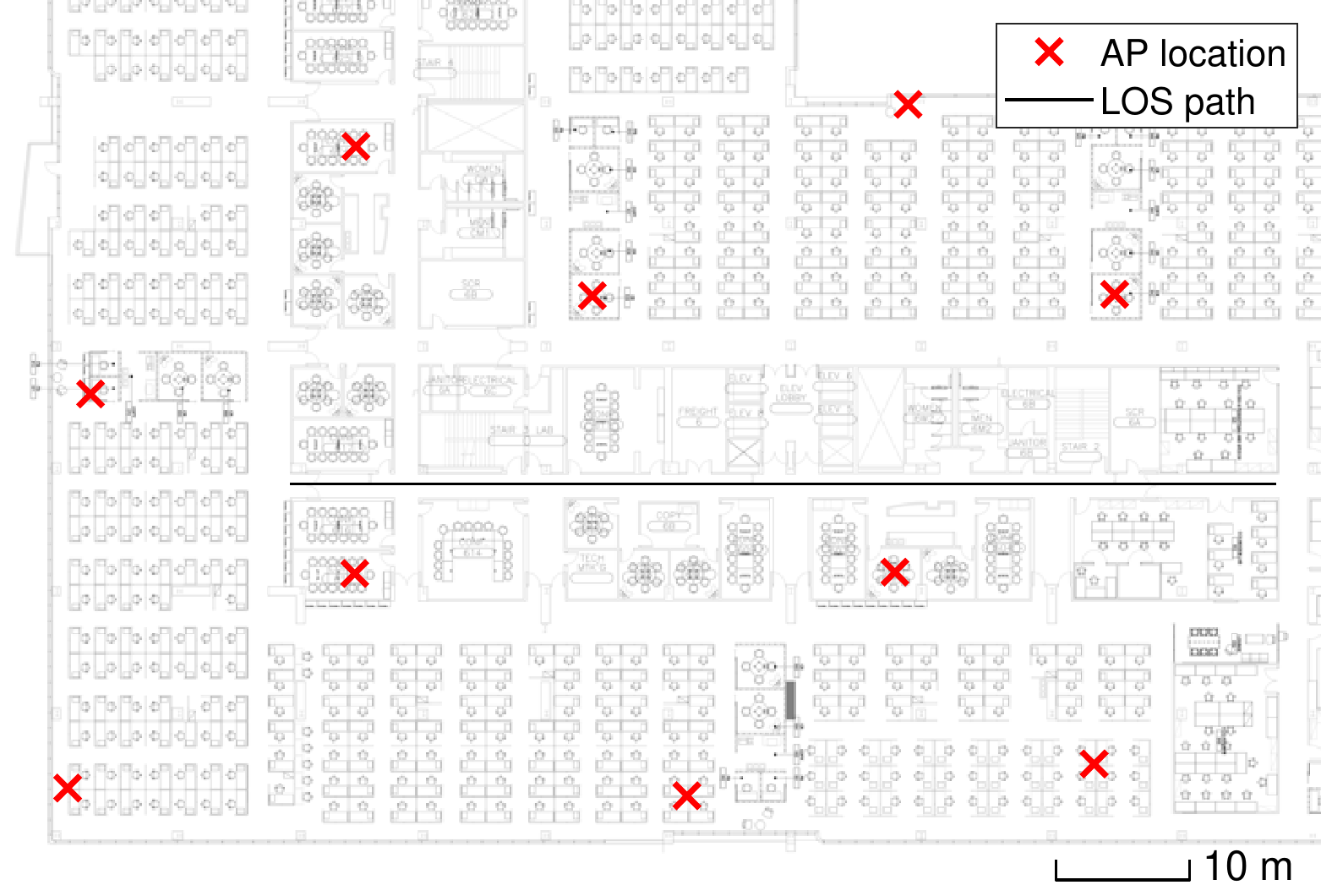}
    \caption{Experiment site with 10 Wi-Fi APs that support the FTM protocol.}
    \label{fig_exp_site}
\end{figure}

To evaluate the ranging performance, we conducted measurement campaigns in an indoor office environment.
Fig.~\ref{fig_exp_site} depicts the corresponding floor plan, where 10 APs are installed at 85~$\times$~55~m$^2$ area\footnote{The APs are deployed in a sparse manner compared to our previous experiments, wherein 10 APs were placed in a 56~$\times$~37~m$^2$ area~\cite{8839041, 8911824}. In this study, the density of an AP is decreased by half to cover a wider area using the same number of APs.}. Each AP is placed on the cubicle walls or in conference rooms at a height of 1.5~m from the floor.
The APs are equipped with an Intel AC8260 Wi-Fi chipset and operated at 5~GHz frequency using Wi-Fi channel 36 with a carrier frequency of 5180~MHz.
Additionally, a laptop equipped with an Intel AX200 Wi-Fi chipset was used as the mobile device.
The device runs on the Ubuntu 18.04 operating system with the latest \emph{iwlwifi} driver (core56).
Thus, the ranging results using the FTM protocol were obtained by executing the \emph{iw} command.

\begin{figure}
    \centering
    \subfloat[]{\includegraphics[width=0.225\textwidth]{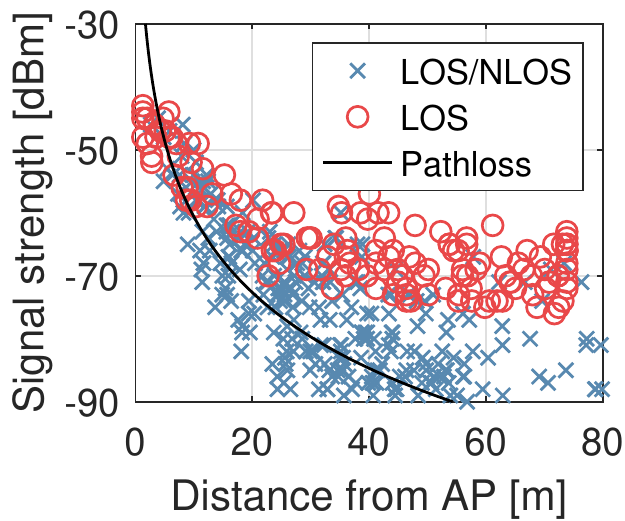}}\hfil\hfil
    \subfloat[]{\includegraphics[width=0.225\textwidth]{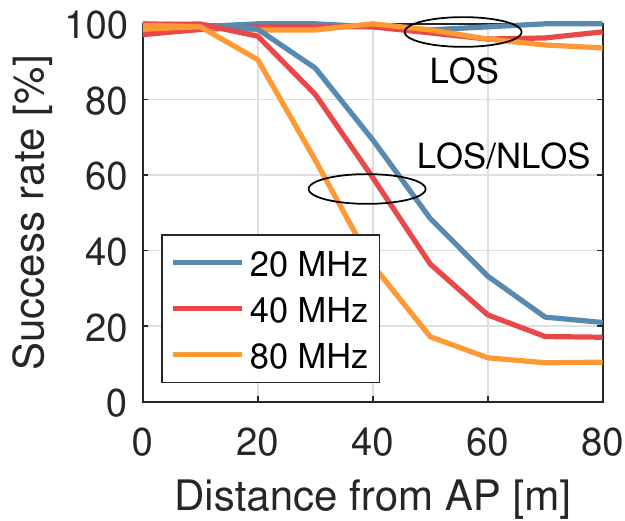}}
    \caption{Ranging results depending on the distance from an AP: (a)~Received signal strength and (b)~success rate of the FTM protocol.}
    \label{fig_ranging_result}
\end{figure}

Initially, we selected 500 different positions in the site and performed the ranging procedure toward all APs at each position.
The measured data collected in this manner were labeled as \emph{LOS/NLOS}, which identifies whether the propagation channel between an AP and the device was of LOS or NLOS condition.
In addition, we installed an AP temporarily at the left end of the LOS path presented in Fig.~\ref{fig_exp_site} to collect the ranging results at several positions on the path.
These data were labeled as \emph{LOS}.
We collected the ranging results at each position considering all supported bandwidths, namely 20, 40, and 80~MHz, and set the burst mode with $B=8$ to obtain more reliable distance measurements under the multipath propagation environment.

Fig.~\ref{fig_ranging_result}(a) demonstrates that the signal strength of the \emph{LOS/NLOS} data decays rapidly with distance than that of the \emph{LOS} data.
The solid black line indicates the path loss curve, which predicts the distance corresponding to the RSS measurement $p$ as
\begin{equation} \label{3A1}
    \hat{d}_{PL}(p) = d_0 10^{\frac{p_0 - p}{10 \eta}},
\end{equation}
where $p_0$ indicates the RSS at a reference distance denoted by $d_0$, and $\eta$ represents the path loss exponent.
The parameters for the path loss curve depicted in Fig.~\ref{fig_ranging_result}(a) are selected using the \emph{LOS/NLOS} data to minimize the mean squared error (MSE) between the estimated and true distances.
The selected parameters are $d_0=1$~m, $p_0=-20.6$~dBm, and $\eta=4$. 
Fig.~\ref{fig_ranging_result}(b) indicates that the success rate of the FTM protocol in the case of \emph{LOS} data is always higher than 95\% irrespective of the distance.
Conversely, the success rate in the case of \emph{LOS/NLOS} data tends to decrease with distance as the RSS approaches the noise floor with increasing distance.

\begin{figure}
    \centering
    \subfloat[]{\includegraphics[width=0.225\textwidth]{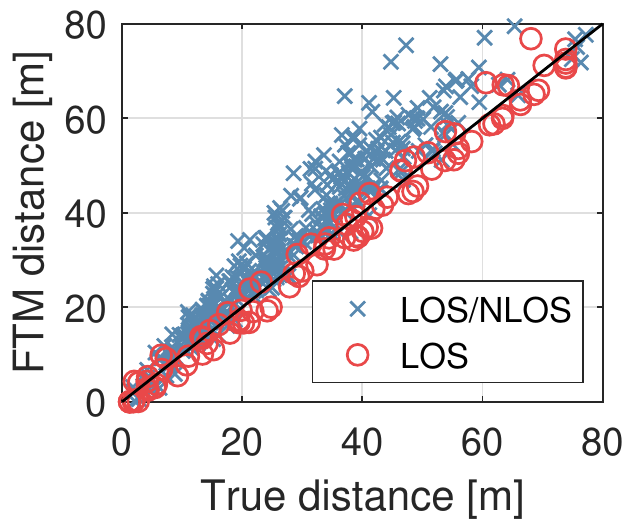}}\hfil\hfil
    \setcounter{subfigure}{3}
    \subfloat[]{\includegraphics[width=0.225\textwidth]{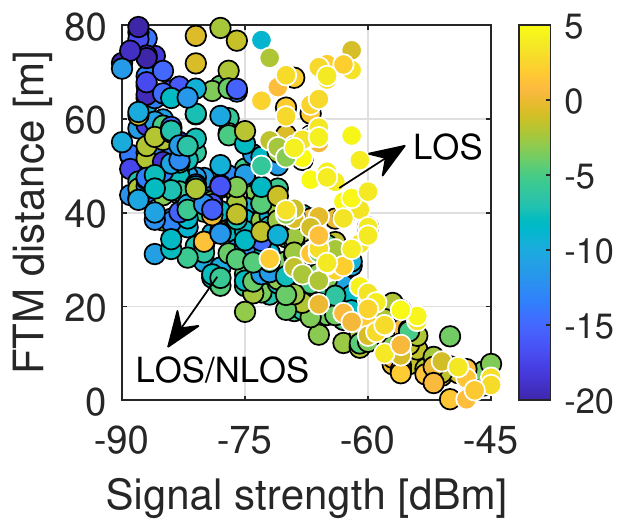}}\\
    \setcounter{subfigure}{1}
    \subfloat[]{\includegraphics[width=0.225\textwidth]{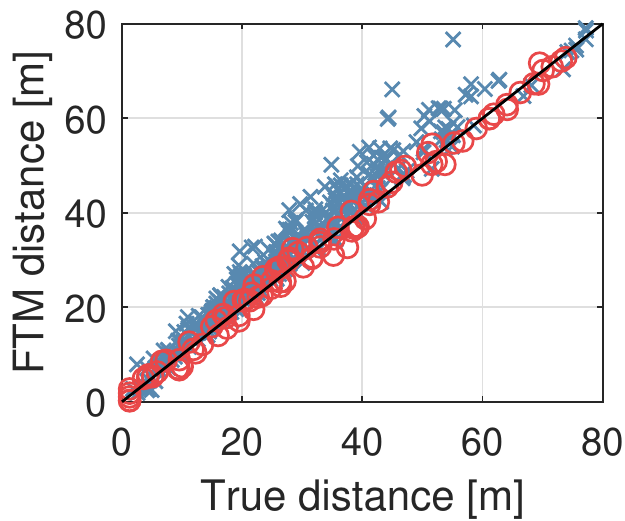}}\hfil\hfil
    \setcounter{subfigure}{4}
    \subfloat[]{\includegraphics[width=0.225\textwidth]{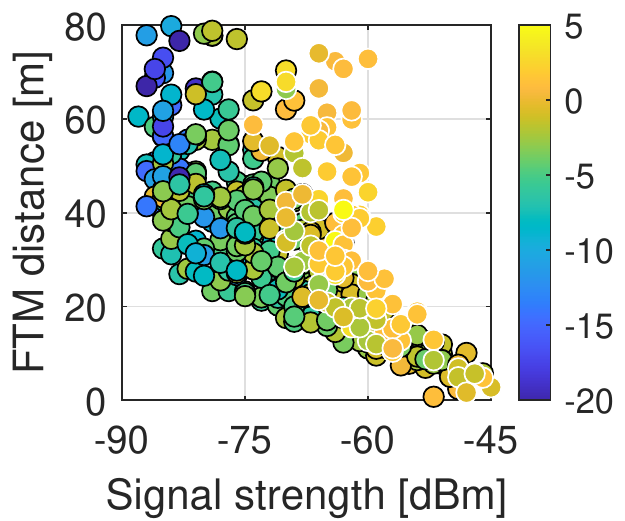}}\\
    \setcounter{subfigure}{2}
    \subfloat[]{\includegraphics[width=0.225\textwidth]{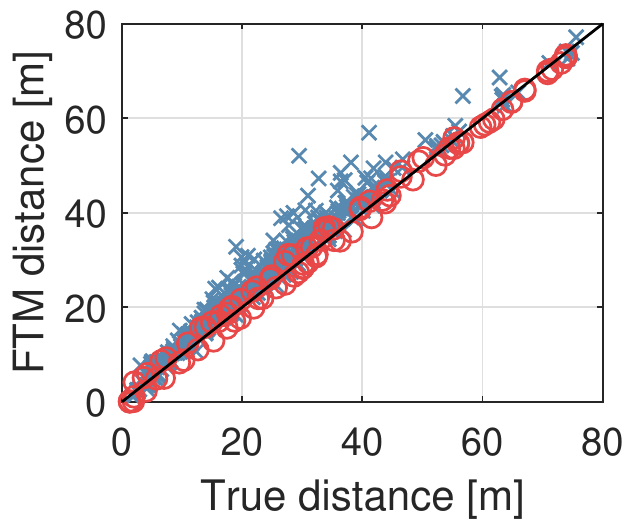}}\hfil\hfil
    \setcounter{subfigure}{5}
    \subfloat[]{\includegraphics[width=0.225\textwidth]{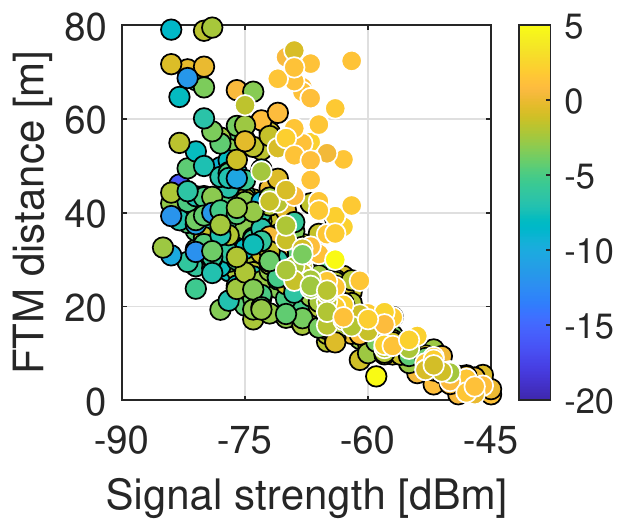}}
    \caption{Relationship between the estimated and true distance for bandwidths of (a) 20~MHz, (b) 40~MHz, and (c) 80~MHz. The pattern of ranging error for bandwidths of (d) 20~MHz, (e) 40~MHz, and (f) 80~MHz.}
    \label{fig_ranging_result_bw}
\end{figure}

Fig.~\ref{fig_ranging_result_bw}(a), (b), and (c) illustrate the relationship between the ground truth distance and raw distance measurement obtained using the FTM protocol at different bandwidths.
The distance measurements of the \emph{LOS} data are concentrated at the solid black lines irrespective of the bandwidths, indicating an ideal scenario.
Conversely, the distance measurements of the \emph{LOS/NLOS} data tend to generate longer distance values than the true distance, which result in inaccurate positioning results.
Fig.~\ref{fig_ranging_result_bw}(d), (e), and (f) illustrate the relationship between the measured distance and RSS, wherein the color of each point represents the difference between the true and measured distances.
The points with white and black edges indicate the \emph{LOS} and \emph{LOS/NLOS} data, respectively.

\section{Enhanced Wi-Fi RTT Ranging Using A Neural Network}

\subsection{Ranging Module}

As indicated in Fig.~\ref{fig_ranging_result_bw}, the ranging error of the FTM protocol has a specific pattern depending on the measured distance and RSS.
To learn such patterns efficiently, we deploy an NN that produces refined ranging results using raw measurements.
The input layer includes all outputs from the FTM protocol as
\begin{equation} \label{4a1}
    \mathbf x = [d_{ftm}, s_{ftm}, p_{ftm}]^T,
\end{equation}
where $d_{ftm}$, $s_{ftm}$, and $p_{ftm}$ represent the distance, standard deviation, and RSS measurements, respectively.

Because the input layer is a vector of three values, fully-connected (FC) layers are applied in this study. 
We deploy $L$ hidden layers between the input and output layers, wherein the $l$-th layer has $D_l$ nodes.
The activation of the $l$-th layer is expressed as
\begin{equation} \label{4a2}
    \mathbf h_l = \sigma\left(\mathbf W_l\mathbf h_{l-1} + \mathbf b_l\right),~1\leq l \leq L,
\end{equation}
where $\mathbf W_l$ represents a $D_l \times D_{l-1}$ weight matrix between the $l-1$ and $l$-th layers, $\mathbf b_l$ denotes a $D_l \times 1$ bias vector, and $\sigma(\cdot)$ is the sigmoid activation function.
Additionally, $\mathbf h_0$ denotes the input layer considered by equation~(\ref{4a2}).

The final hidden layer generates two scalar outputs that indicate the estimated distance and the corresponding expected standard deviation.
These values are computed as
\begin{align} \label{4a3}
    \hat{d} &= \bar{d} \sigma(\mathbf W_d \mathbf h_L + b_d),\nonumber\\
    \hat{s} &= \bar{s} \sigma(\mathbf W_s \mathbf h_L + b_s),
\end{align}
where the symbols denoted by $\mathbf W$ are $1\times D_L$ weight matrices and those denoted by $b$ represent scalar biases.
Because the sigmoid function produces a value in the range 0 to 1, each output is upper bounded by $\bar{d}$ and $\bar{s}$, respectively.

\subsection{Learning Objective}

The aforementioned ranging module is used to obtain enhanced ranging results from each AP.
The relationship between the input and output of the ranging module is expressed as a parameterized function as
\begin{equation} \label{4b1}
  \mathbf g(\mathbf x; \Theta) = [\hat{d}, \hat{s}]^T,
\end{equation}
where $\Theta$ represents the set of all trainable parameters in the FC layers, such as the weight matrices and biases.
Based on this expression, the list of ranging results at time step $k$ is represented as
\begin{equation} \label{4b2}
    \mathcal R^{(k)}(\Theta) = \{\mathbf g(\mathbf x_n^{(k)}; \Theta)\}_{n=1}^N,
\end{equation}
where $\mathbf x_n^{(k)}$ denotes the input vector comprising raw measurements from the $n$-th AP at time step $k$.

As the ranging results depend on each parameter in the set $\Theta$, the estimated Wi-Fi trajectory obtained using equation~(\ref{2B1}) is expressed as a function of the set as
\begin{equation} \label{4b3}
    \hat{\mathbf z}^{(k)}(\Theta) = f(\hat{\mathbf z}^{(k-1)}(\Theta), \mathcal R^{(k)}(\Theta)).
\end{equation}
This implies that the shape of the estimated Wi-Fi trajectory widely varies depending on the set of trainable parameters in the ranging module.
Therefore, the objective of this paper is to optimize every trainable parameter so that the Wi-Fi trajectory becomes accurate.
In addition, to train the ranging module without collecting ground truth data, we utilize a reference trajectory provided from sensors and train the ranging module by comparing the shape of the Wi-Fi trajectory and the reference trajectory.

\subsection{Ranging Module Training with Sensor-Aided Learning}

In this study, the reference trajectory is obtained using equation~(\ref{2b1}) based on the PDR method.
To synchronize the time between the Wi-Fi and PDR trajectories, we define $\mathbf p^{(k)} = \mathbf p(t_k)$, where $t_k$ denotes the sensor time during the $k$-th Wi-Fi ranging procedure. For ease of exposition, we consider a single dataset in which the Wi-Fi ranging procedure is performed $K$ times.

\begin{figure}
    \centering
    \subfloat[]{\includegraphics[width=0.4\textwidth]{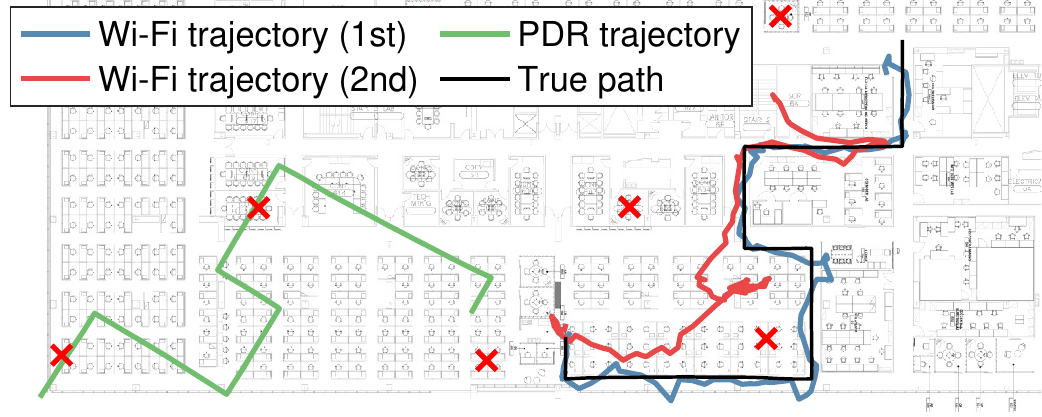}}\\
    \subfloat[]{\includegraphics[width=0.225\textwidth]{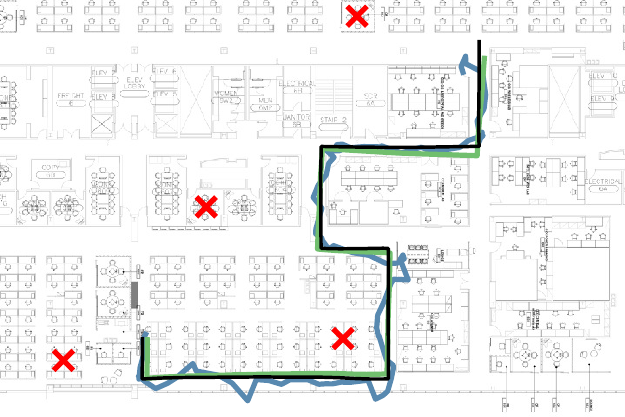}}\hfil\hfil 
    \subfloat[]{\includegraphics[width=0.225\textwidth]{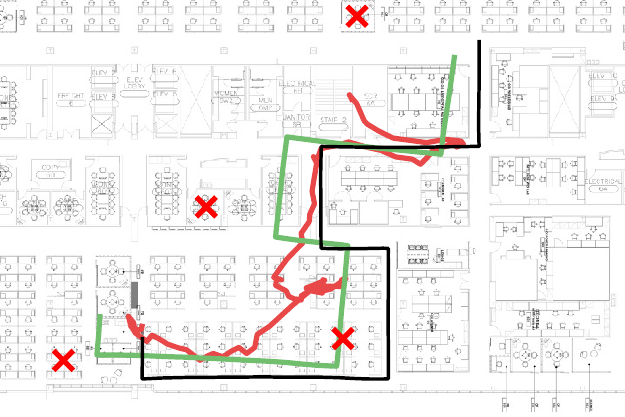}}
    \caption{Overview of the sensor-aided learning technique. (a) Estimated trajectories using Wi-Fi ranging results with two different sets of parameters and raw PDR trajectory are presented. The PDR trajectory is transformed close to (b) the first Wi-Fi trajectory and (c) the second Wi-Fi trajectory to determine the similarity in shape.}
    \label{fig_sen_overview}
\end{figure}

Fig.~\ref{fig_sen_overview} illustrates an overview of the sensor-aided learning technique. Fig.~\ref{fig_sen_overview}(a) depicts two estimated trajectories based on Wi-Fi ranging results with different sets of parameters. Although the shape of the Wi-Fi trajectory varies depending on the set of parameters, both trajectories are located close to the true path as the locations of nearby APs are used in the positioning stage.
Conversely, the trajectory obtained according to the PDR method begins from an arbitrary initial position, and the direction is not aligned with the test path as $\phi_{ref}$ is determined randomly. 
Nevertheless, the shape of the PDR trajectory is nearly similar to that of the test path.
Therefore, comparing the shapes of the Wi-Fi and PDR trajectories is equivalent to comparing the shapes of the the Wi-Fi trajectory and the test path.
Thus, the PDR trajectory is rotated suitably and shifted close to the Wi-Fi trajectory to compare the shapes, as depicted in Fig.~\ref{fig_sen_overview}(b) and (c).

The cost function that computes the MSE between the Wi-Fi and PDR trajectories after performing an optimal transformation operation is given by~\cite{choi2020sensoraided}
\begin{align} \label{4b4}
    \mathcal{L}(\Theta) = &\sum_k \lVert \hat{\mathbf z}^{(k)}(\Theta)\rVert^2 + \sum_k \lVert {\mathbf p}^{(k)}\rVert^2- 2\sqrt{\Gamma^2 + \tilde{\Gamma}^2}\nonumber\\
    & +\frac{\lVert\sum_k \hat{\mathbf z}^{(k)}(\Theta)\rVert + \lVert \sum_k \mathbf p^{(k)}\rVert}{K},
\end{align}
where $\Gamma$ and $\tilde{\Gamma}$ are associated with the two trajectories as
\begin{align}  \label{4b5}
\Gamma &= \frac{(\sum_k \hat{\mathbf{z}}^{(k)}(\Theta))^T(\sum_k\mathbf p^{(k)})}{K} - \sum_k (\hat{\mathbf z}^{(k)}(\Theta))^T\mathbf p^{(k)},\nonumber\\
\tilde{\Gamma} &= \frac{(\sum_k \hat{\mathbf{z}}^{(k)}(\Theta))^T\tilde{\mathbf I}(\sum_k\mathbf p^{(k)})}{K} - \sum_k (\hat{\mathbf z}^{(k)}(\Theta))^T\tilde{\mathbf I}\mathbf p^{(k)}.
\end{align}
Here, $\tilde{\mathbf I}$ denotes a 2$\times$2 anti-diagonal matrix with $[\tilde{\mathbf I}]_{(1,2)}=-1$ and $[\tilde{\mathbf I}]_{(2,1)}=1$.
All summation operations in equations~(\ref{4b4}) and (\ref{4b5}) are determined for $1\leq k \leq K$.
Using this cost function, the ranging module is trained by following the sensor-aided learning technique, which is outlined in Algorithm~1.

\begin{algorithm}
\caption{Sensor-Aided Learning Procedure}
\begin{algorithmic}[1]
\renewcommand{\algorithmicrequire}{\textbf{Input:}}
\renewcommand{\algorithmicensure}{\textbf{Output:}}
\REQUIRE Training and validation datasets, each consists of raw FTM measurements $\mathbf{x}_n^{(k)}$ and PDR trajectory $\mathbf p^{(k)}$ for $k=1,...,K$ and $n=1,...,N$
\ENSURE Set of trained parameters $\Theta^*$\\
\STATE \emph{(Initialization)} Randomly initialize $\forall \theta \in \Theta$, set $\mathcal L_{val} \leftarrow\infty$
	\FOR{each epoch}
	\STATE \emph{(With training dataset)}
    \STATE Compute ranging results $\mathbf g(\mathbf x_n^{(k)}; \Theta)$ for $\forall k, n$
    \STATE Compute Wi-Fi trajectory using equation~(\ref{2B1}) \\
    \STATE Compute cost $\mathcal L(\Theta)$ using equation~(\ref{4b4})\\
    \STATE Derive derivatives $\frac{\partial \mathcal L (\Theta)}{\partial \theta}$ for $\forall \theta\in\Theta$\\
    \STATE Update parameters as $\theta \leftarrow \theta - \mu\frac{\partial \mathcal L (\Theta)}{\partial \theta}$  for $\forall \theta\in\Theta$
    \STATE \emph{(With validation dataset)}
    \STATE Compute cost $\mathcal L(\Theta)$ by following steps 4-6
    \IF {$\mathcal L(\Theta) < \mathcal L_{val}$}
    \STATE $\Theta^*\leftarrow\Theta$, $\mathcal L_{val}\leftarrow \mathcal L(\Theta)$
    \ENDIF
	\ENDFOR
\end{algorithmic}
\end{algorithm}
In this procedure, $\mu$ used in step~8 represents the learning rate. In case of multiple training or validation datasets used in Algorithm~1, we compute the cost of each dataset by following steps 4-6 and compute the consolidated cost by taking the sum of each cost.

\section{Performance Evaluation}

The ranging and positioning performances were evaluated at the site depicted in Fig.~\ref{fig_exp_site}.
To simulate multiple users accessing location services, six participants moved around the indoor area freely for 5 min each.
The Wi-Fi ranging procedure was performed for every 1~s throughout the experiment, and the prameters for the FTM protocol were set the same as those used in Section~III.

In addition to Wi-Fi data, accelerometer and gyroscope readings were collected at a sampling rate of 100~Hz to obtain reference trajectories using the PDR method.
The collected data were split into multiple datasets, each corresponding to 100 Wi-Fi time steps (i.e., 100~s).
Herein, 70\% of the datasets were used for training and the remaining for validation purposes.
In addition, the test data were collected separately by moving along a predefined test path of length 520~m.
The major experimental parameters are summarized in Table~\ref{tab_exp_parameters}.

\begin{table}
    \renewcommand{\arraystretch}{1.25}
    \caption{Experimental parameters}
    \centering
    \begin{tabular}{cc}
    \hline
    {\bf Parameters} & {\bf Values}\\
    \hline
    Experiment site & Office with 85~$\times$~55~m$^2$ area\\
    Number of APs & 10\\
    AP model & WiLD from Compulab\\
    Mobile device model & Laptop running on Ubuntu 18.04\\
    Wi-Fi chipset (AP) & Intel Wireless AC8260\\
    Wi-Fi chipset (mobile) & Intel Wi-Fi 6 AX200\\
    Antenna height & 1.5~m (both AP and device)\\
    Wi-Fi center frequency & 5180~MHz (Wi-Fi channel 36)\\
    FTM burst mode & Enabled with $B=8$\\
    Bandwidth of FTM signals  & 20, 40, and 80~MHz\\
    Ranging procedure frequency & 1 Hz\\
    Sensor model & BNO055 from Bosch Sensortec\\
    Collected sensor data & Accelerometer, gyroscope readings\\
    Sensor sampling rate & 100 Hz\\
    Ranging module & Two FC layers each with 100 nodes\\
    Upper bound of ranging outputs & $\bar{d} = 100$~m, $\bar{s} = 10$~m\\
    Number of participants & 6\\
    Collected training data & Wi-Fi and sensor data for 30~minutes\\
    Data split ratio    & 70~\% (training), 30~\% (validation)\\
    Test path length    & 520~m\\
    \hline
    \end{tabular}
    \label{tab_exp_parameters}
\end{table}

\subsection{Training Stage}

We deployed two hidden layers, each with 100 hidden nodes, for the ranging module implemented using FC layers.
The upper bounds of the distance and standard deviation output were given by $\bar{d}=100$ and $\bar{s}=10$~m, respectively.
The trajectory of the device is obtained using ranging results from up to 5 nearby APs.
Additionally, the similarity between the Wi-Fi and PDR trajectories is evaluated to optimize all the trainable parameters in the ranging module.

\begin{figure}
    \centering
    \subfloat[]{\includegraphics[width=0.225\textwidth]{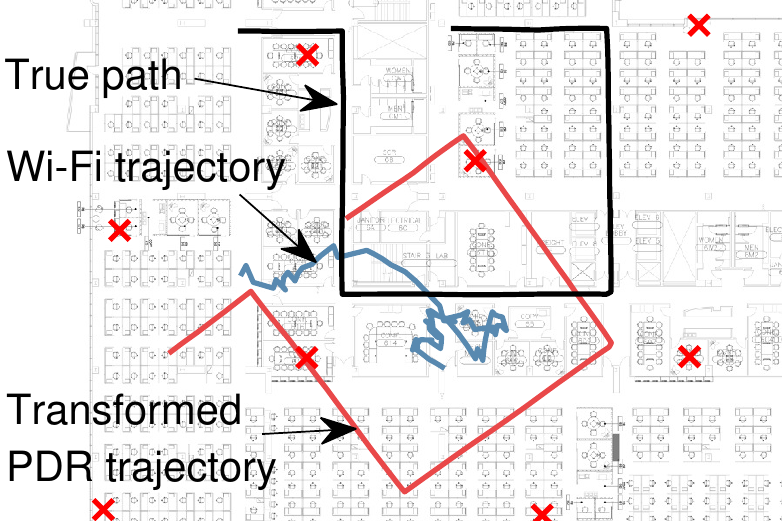}}\hfil\hfil
    \subfloat[]{\includegraphics[width=0.225\textwidth]{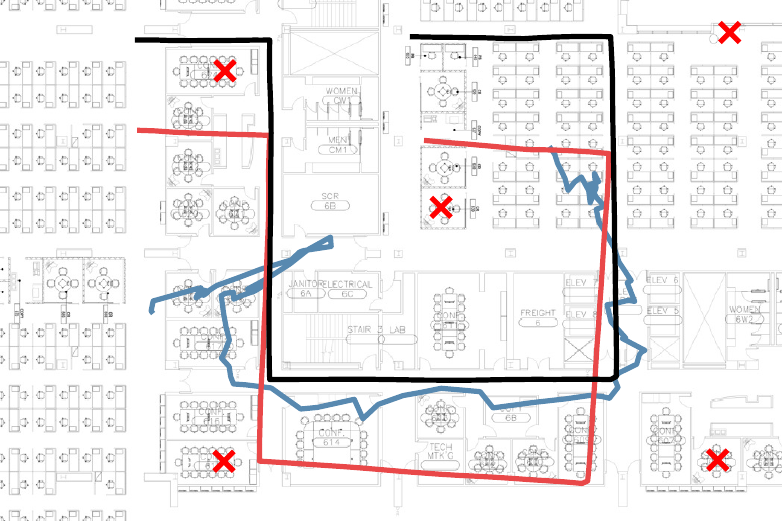}}\hfil\hfil
    \subfloat[]{\includegraphics[width=0.225\textwidth]{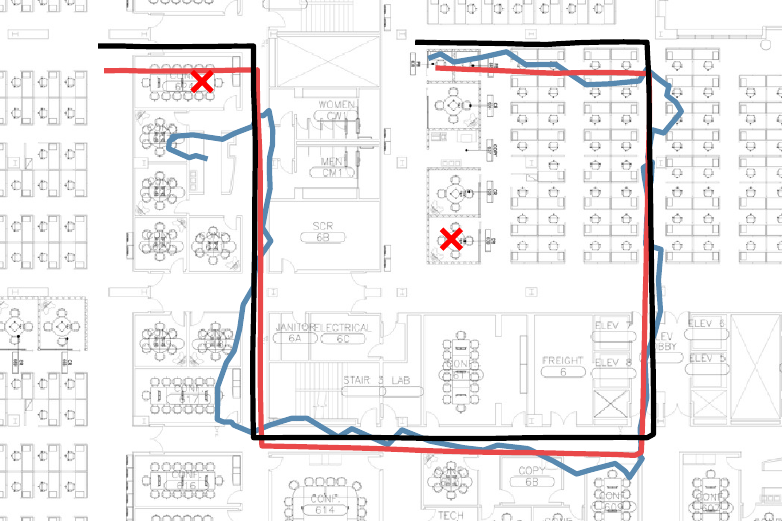}}\hfil\hfil
    \subfloat[]{\includegraphics[width=0.225\textwidth]{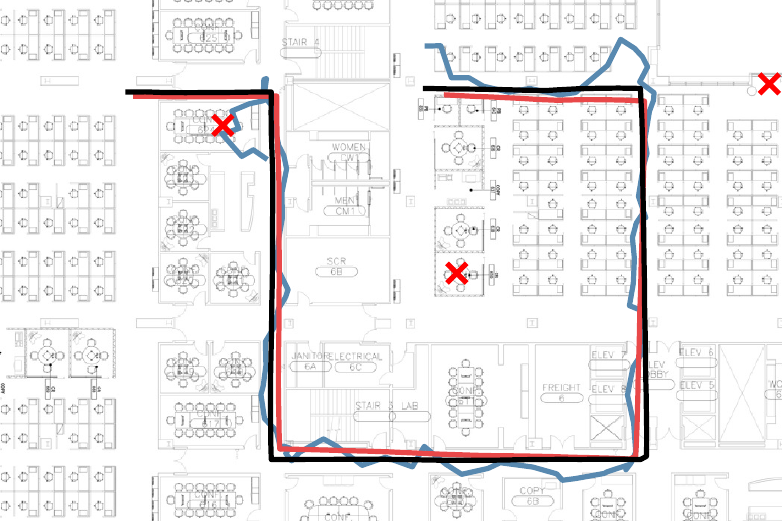}}
    \caption{Estimated trajectory of the device using the output of the ranging module obtained at different training epochs:  (a) epoch 1, (b) epoch 20, (c) epoch 100, and (d) epoch 200. The PDR trajectory is transformed close to the Wi-Fi trajectory to determine the similarity between the two trajectories.}
    \label{fig_nn_trj}
\end{figure}

Fig.~\ref{fig_nn_trj} depicts the true path, Wi-Fi, and PDR trajectories observed at a specific dataset. 
The PDR trajectory indicates the transformed version, which is used to determine the MSE from the Wi-Fi trajectory.
As all training parameters are randomly initialized at the beginning of the training process, the ranging module produces arbitrary outputs regardless of the input.
Consequently, an inaccurately estimated trajectory is obtained based on these ranging results as shown in Fig.~\ref{fig_nn_trj}(a), wherein the shape varies substantially from that of the PDR trajectory.

However, as the training process proceeds, the parameters in the ranging module are adjusted to ensure gradual similarity between the shapes of the WiFi and PDR trajectories, as shown in Fig.~\ref{fig_nn_trj}(b)--(d).
Particularly, Fig.~\ref{fig_nn_trj}(d) indicates that the shape of the Wi-Fi and PDR trajectories become nearly similar after 200 training epochs, and the transformed PDR trajectory overlaps with the test path completely.
Therefore, the MSE between the Wi-Fi and transformed PDR trajectories is equivalent to that between the Wi-Fi trajectory and the true path.
Thus, the ranging module is trained without collecting ground truth trajectory.

\begin{figure}
    \centering
    \subfloat[]{\includegraphics[width=0.225\textwidth]{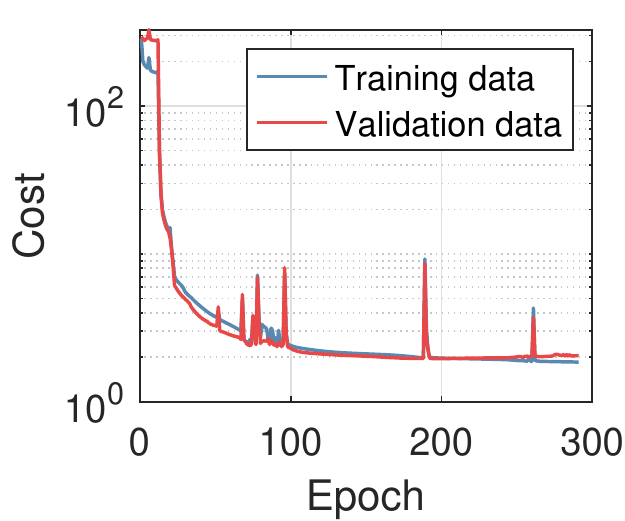}}\hfil\hfil
    \subfloat[]{\includegraphics[width=0.225\textwidth]{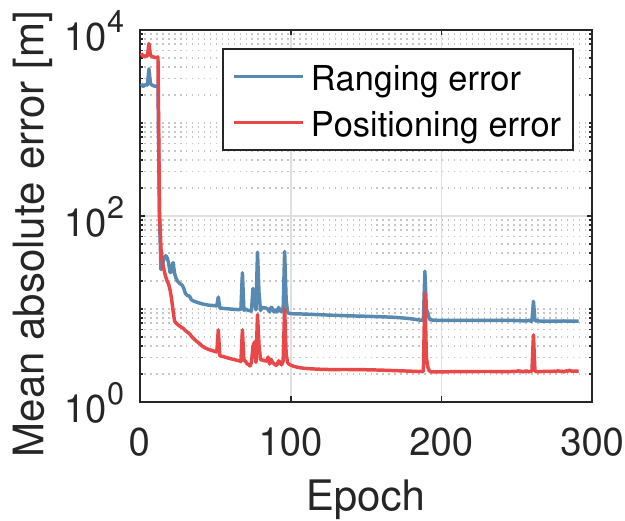}}\hfil\hfil
    \subfloat[]{\includegraphics[width=0.225\textwidth]{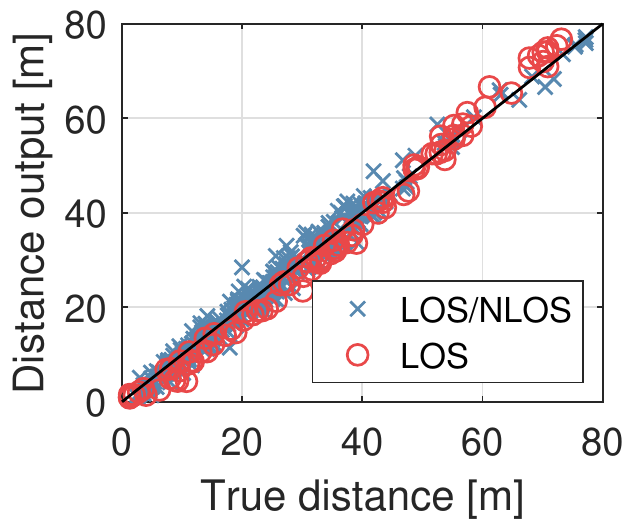}}\hfil\hfil
    \subfloat[]{\includegraphics[width=0.225\textwidth]{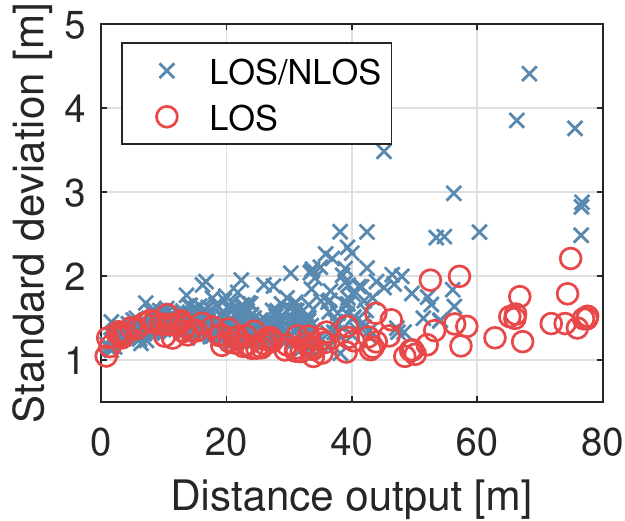}}
    \caption{Training results with the data collected using 40~MHz bandwidth: (a) Training and validation cost, (b) ranging and positioning error, (c) relationship between true distance and distance output of the ranging module, and (d) relationship between the two outputs of the ranging module.}
    \label{fig_nn_cost}
\end{figure}

Fig.~\ref{fig_nn_cost} illustrates the training results for the 40~MHz bandwidth scenario.
Fig.~\ref{fig_nn_cost}(a) verifies that both the training and validation costs decrease with the training epoch. Consequently, the ranging and positioning errors with respect to the test data decrease with the epoch, as shown in Fig.~\ref{fig_nn_cost}(b).
Additionally, the ranging performance was evaluated considering the labeled data collected in Section~III.
Fig.~\ref{fig_nn_cost}(c) depicts the relationship between the true distance and estimated distance obtained from the ranging module.
Herein, the points are located closer to the solid black line compared to those in Fig.~\ref{fig_ranging_result_bw}(b).
Finally, Fig.~\ref{fig_nn_cost}(d) depicts the relationship between the two outputs of the ranging module.
According to the figure, the ranging module was trained to generate a lower standard deviation for the \emph{LOS} data, meaning that the EKF-based positioning module more relies on ranging results obtained in LOS conditions.

\begin{figure}
    \centering
    \includegraphics[width=0.35\textwidth]{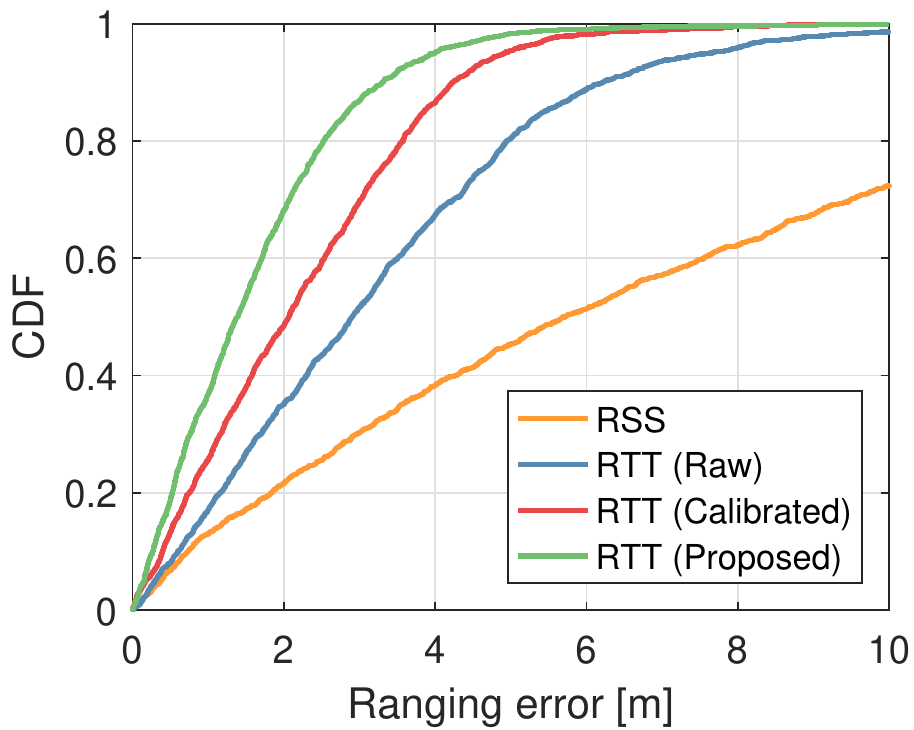}
    \caption{Cumulative density function of ranging error (40~MHz).}
    \label{fig_cdf_ranging}
\end{figure}

Fig.~\ref{fig_cdf_ranging} illustrates the cumulative density function (CDF) of the ranging error. 
To compare the performances, we evaluated the ranging performance using RSS measurements based on the path loss curve obtained from equation~(\ref{3A1}).
In addition, a well-calibrated scenario was considered where a bias $\delta$ in the raw distance measurements is calibrated as
\begin{equation}\label{5a1}
    \hat{d}_{cal}(d_{ftm}) = \max(0, d_{ftm} - \delta).
\end{equation}
The max operator is applied to ensure that the calibrated distance is always non-negative. 
An optimal bias is selected using the \emph{LOS/NLOS} data to minimize the MSE of ranging error.
The selected biases are given by $\delta = 6.6$, $4.4$, and $3.4$~m for the bandwidths of 20, 40, and 80~MHz, respectively. 
The CDF indicates that ranging determined using RSS produces inaccurate results compared to other ranging results.
As the bias in the raw distance measurements obtained from the FTM protocol is calibrated, the ranging results can be obtained with enhanced precision.
Furthermore, the proposed ranging module produces the most accurate results by learning the patterns observed in the raw measurements.

\begin{figure}
    \centering
    \subfloat[]{\includegraphics[width=0.225\textwidth]{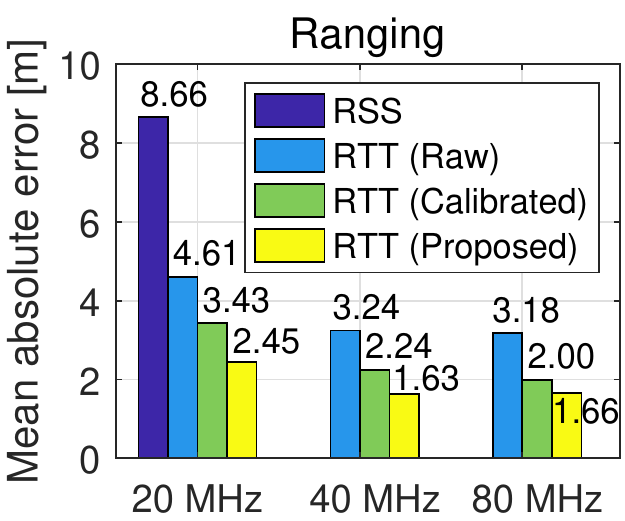}}\hfil\hfil
    \subfloat[]{\includegraphics[width=0.225\textwidth]{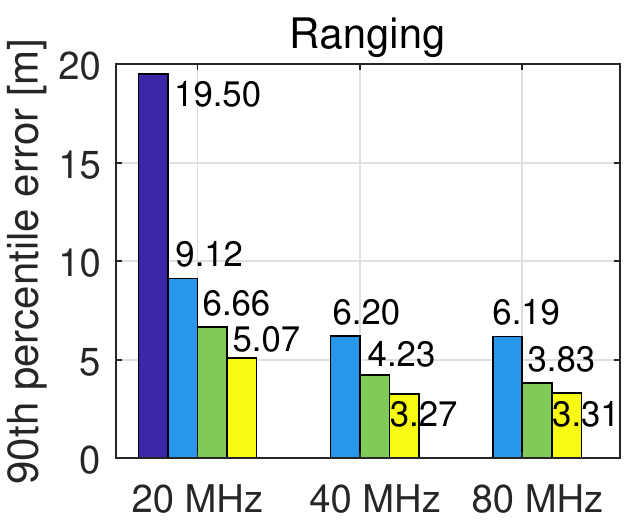}}
    \caption{Ranging performance depending on the bandwidth: (a) Mean absolute error and (b) 90th percentile error.}
    \label{fig_bar_ranging}
\end{figure}

Fig.~\ref{fig_bar_ranging}(a) depicts the mean absolute error (MAE) of the ranging results.
The proposed method reduces the error of the raw distance measurement by 47--50\% and even outperforms the well-calibrated scenario, which requires to collected ground truth distances.
This is because the NN autonomously identifies channel conditions from the raw distance, standard deviation, and RSS measurements to produce enhanced ranging results.
Consequently, the proposed method reduces the MAE by 17--29\% in case of well-calibrated ranging scenario.
Additionally, the 90th percentile error exhibits a similar trend, as shown in Fig.~\ref{fig_bar_ranging}(b).
The proposed ranging method reduces the error by 44--47\% and 14--24\% from the raw distance measurement and the well-calibrated ranging scenario, respectively.

\subsection{Online Stage}

In this subsection, we evaluate the positioning performance considering two scenarios. The first scenario involves only Wi-Fi ranging in the positioning process, whereas the second scenario uses sensors along with Wi-Fi ranging.
When sensors are used in location services, sensor data can be collected and utilized to train the ranging module based on the sensor-aided learning framework.
To estimate the trajectory of the device using sensors, we followed the latest EKF procedure presented in~\cite{choi2020sensoraided}.
Initially, we evaluate the performance considering ranging results obtained with a bandwidth of 40~MHz.

\begin{figure}
    \centering
    \subfloat[]{\includegraphics[width=0.4\textwidth]{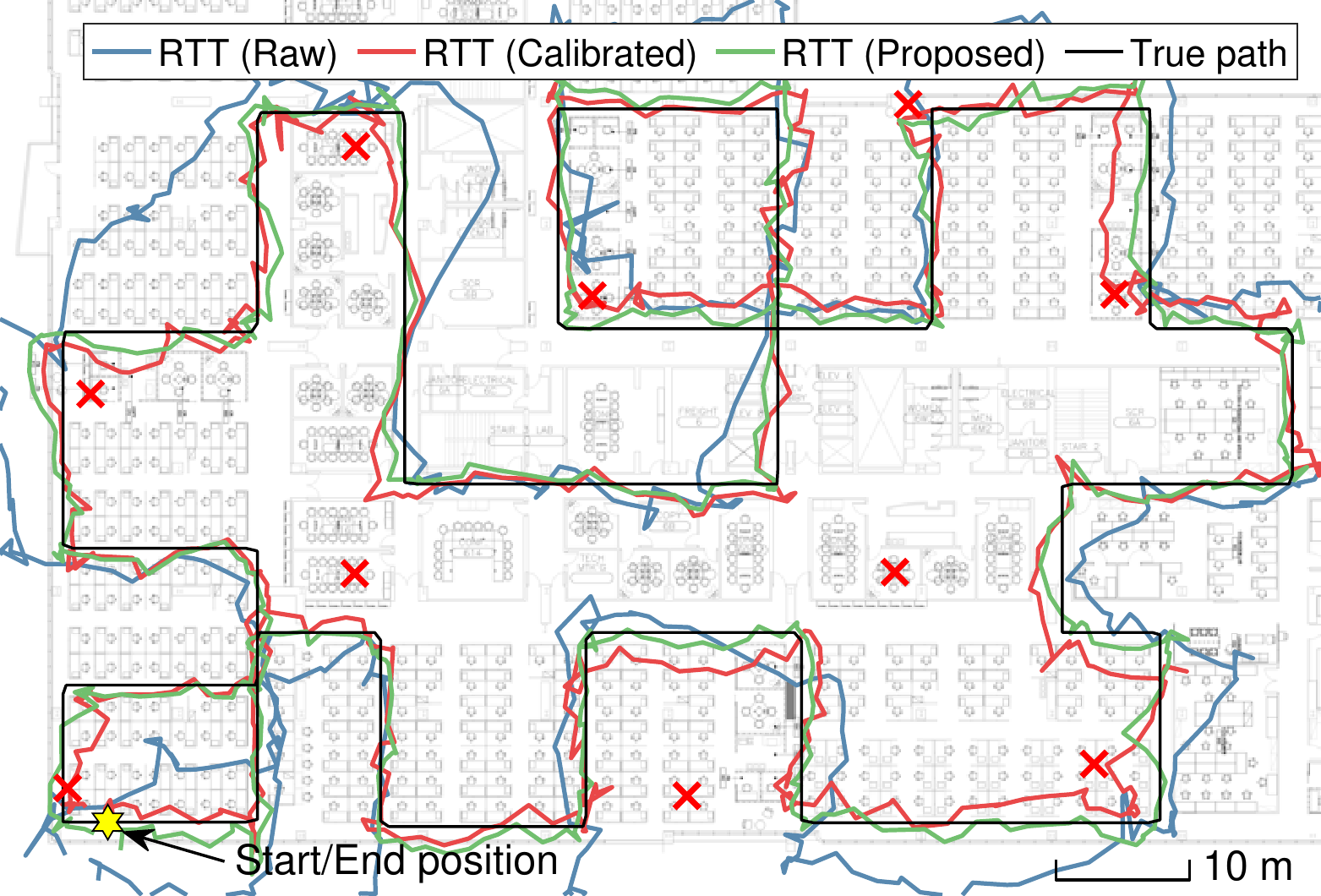}}\\
    \subfloat[]{\includegraphics[width=0.4\textwidth]{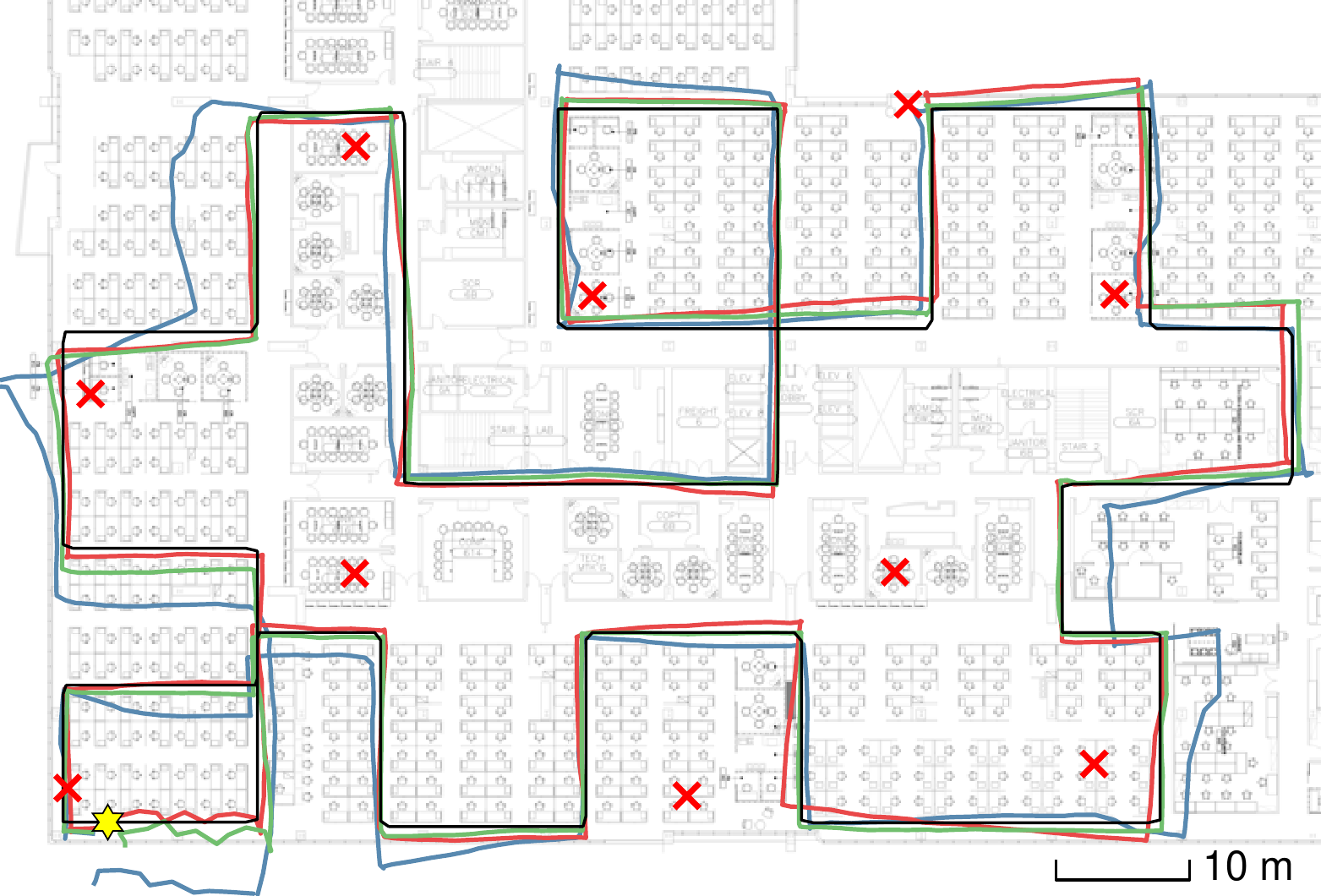}}
    \caption{Estimated trajectory using Wi-Fi ranging with 40~MHz bandwidth: (a) Only ranging results are used and (b) ranging and sensors are used.}
    \label{fig_trj}
\end{figure}

Fig.~\ref{fig_trj}(a) illustrates the estimated trajectory considering only the Wi-Fi ranging results.
As the raw distance measurement obtained from the FTM protocol tends to produce a longer distance value than the true distance, the estimated trajectory generates a substantial error, particularly when APs are concentrated in a specific direction from the position of the device.
For instance, the estimated trajectory near the boundary of the experimental area yields a significant positioning error. 
However, this phenomenon is resolved when the raw distance measurement is calibrated and the estimated trajectory is close to the test path.
In addition, the estimated trajectory obtained using enhanced ranging results is closer to the test path.
Additionally, a smoother trajectory is obtained when the sensors are used in the positioning stage, as shown in Fig.~\ref{fig_trj}(b).

\begin{figure}
    \centering
    \includegraphics[width=0.35\textwidth]{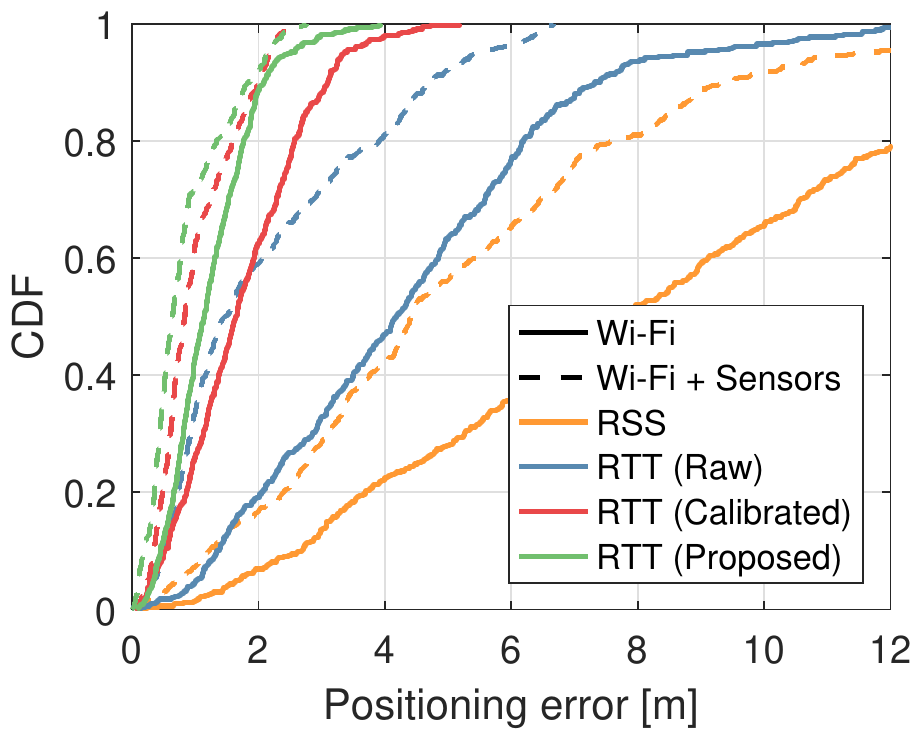}
    \caption{Cumulative density function  of positioning error (40~MHz).}
    \label{fig_cdf_positioning}
\end{figure}

\begin{figure}
    \centering
    \subfloat[]{\includegraphics[width=0.225\textwidth]{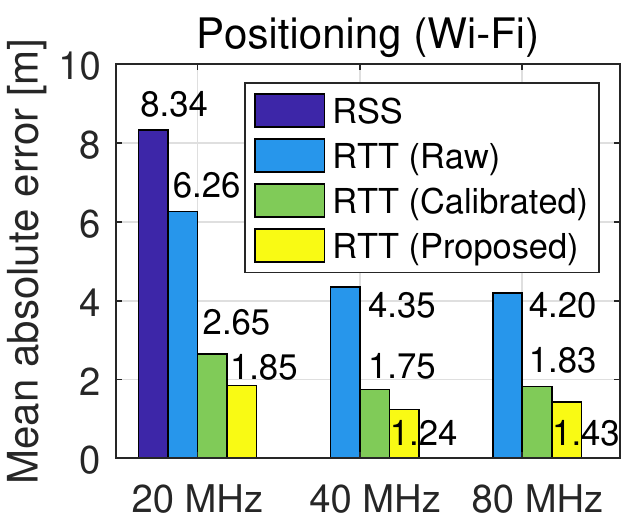}}\hfil\hfil
    \subfloat[]{\includegraphics[width=0.225\textwidth]{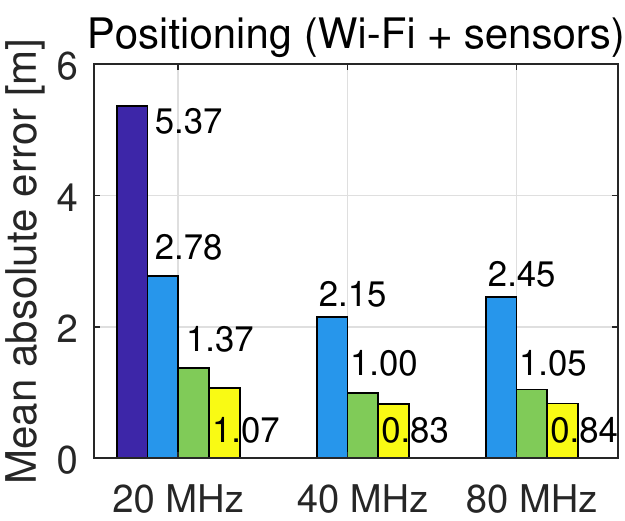}}
    \caption{Mean absolute error of the positioning error: (a) Wi-Fi ranging results are used and (b) sensor are used along with Wi-Fi ranging results.}
    \label{fig_bar_positioning}
\end{figure}

Fig.~\ref{fig_cdf_positioning} depicts the CDF of the positioning error for each ranging scenario. 
The positioning performance depends on the quality of the ranging results as the order of the CDF curves in Fig.~\ref{fig_cdf_positioning} is identical to that observed in Fig.~\ref{fig_cdf_ranging}.
Moreover, the performance can be improved when sensors are used in the positioning stage.
Finally, Fig.~\ref{fig_bar_positioning}(a) and (b) illustrate the MAE of the positioning error with and without sensors, respectively.
Unlike the drastic performance improvement observed between  20 and 40~MHz bandwidth scenarios, no significant performance enhancement is observed between 40 and 80~MHz bandwidth scenarios.
This is because the ranging performances of 40 and 80~MHz bandwidth scenarios in our experimental setup are nearly similar, as shown in Fig.~\ref{fig_bar_ranging}.

Furthermore, as indicated in Fig.~\ref{fig_ranging_result}(b), the success rate of the FTM protocol for the \emph{LOS/NLOS} data tends to decay rapidly with distance for wider bandwidths.
Consequently, the average number of APs involved in the positioning procedure is obtained as 4.6, 4.4, and 3.6 for bandwidths of 20, 40, and 80~MHz, respectively.
Thus, the positioning performance of the 80~MHz bandwidth scenario may be worse than that of the 40~MHz bandwidth scenario occasionally.
Furthermore, the positioning performance generates accurate results using the proposed ranging module despite the absence of sensors. Therefore, location services can decide whether to use sensors based on the required location accuracy, battery life of the device, and other requirements.

\section{Conclusion}

In this paper, we evaluated ranging and positioning performances using the FTM protocol.
Although the FTM protocol is able to potentially determine precise ranging between Wi-Fi devices, a calibration process was essential to remove a bias in raw distance measurements.
Furthermore, the experimental results indicated that different measurement patterns are observed in LOS and NLOS conditions.
The proposed NN-based ranging module autonomously learned such patterns using unlabeled sensor data naturally accumulated when users access location services and thus produced enhanced ranging results.
Consequently, accurate positioning results could be obtained.
Even without using sensors, the proposed method achieved the average positioning accuracy of 1.24~m for the 40~MHz bandwidth scenario, which may be enough for applications that require a room-level accuracy.
For a sub-meter level accuracy, sensors can be utilized along with the enhanced ranging results.
Similarly, location services can decide whether to use sensors based on the requirements of applications, and again, when sensors are used in location services, the collected sensor data can be used to train more accurate ranging module.

\section*{Acknowledgment}

We would like to thank Yang-Seok Choi and Shilpa Talwar from Intel Labs for supporting this project.





%





\begin{IEEEbiography}[{\includegraphics[width=1in,height=1.25in,clip,keepaspectratio]{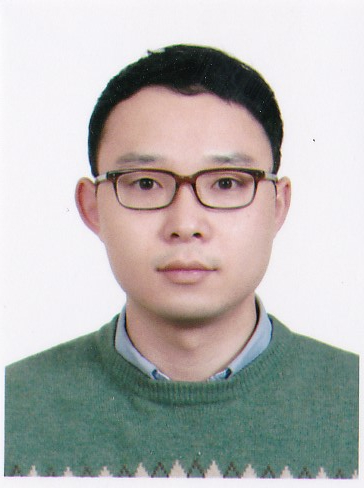}}]{Jeongsik Choi} received the B.S. degree in electrical engineering from Pohang University of Science and Technology (POSTECH), Pohang, South Korea, in 2010, and the M.S. and Ph.D. degrees in electrical engineering and computer science from
Seoul National University, Seoul, South Korea, in 2012 and 2016, respectively. 
From 2016 to 2017, he was a Senior Researcher with the Institute of New Media and Communications, which is a research institute in Seoul National University.
From 2017 to 2021, he was a Research Scientist AI/ML with Intel Labs, Santa Clara, CA, USA. 
Since September 2021, he has been an Assistant Professor with the School of Electronics Engineering, Kyungpook National University, Daegu, South Korea. His research interests include wireless propagation channel
measurement/modeling, sensor fusion and positioning techniques, and he is currently focusing on AI/ML for communication and sensing systems.
\end{IEEEbiography}

%








\end{document}